\documentclass[fleqn,usenatbib]{mnras}

\usepackage[T1]{fontenc}
\usepackage{ae,aecompl}
\usepackage{lastpage}

\usepackage{graphicx}	
\usepackage{amsmath}	
\usepackage{amssymb}	
\usepackage{courier}
\usepackage{xcolor}

\newcommand\be{\begin{equation}}
\newcommand\ee{\end{equation}}
\usepackage{hyperref}
\hypersetup{colorlinks=true,linkcolor=blue,filecolor=magenta,urlcolor=blue}

\usepackage{pdfpages}
\usepackage[inkscapelatex=false]{svg}

\title[Fuzzy Dark Matter Dynamics]{Fuzzy Dark Matter Dynamics and the Quasiparticle Hypothesis}

\author[B. Zupancic et al.]{
Boris Zupancic,\thanks{E-mail: 18bz20@queensu.ca}
Lawrence M. Widrow,\thanks{E-mail: widrow@queensu.ca}
\\
Department of Physics, Engineering Physics and Astronomy, Queen's University, Kingston, K7L 3X5, Canada
}

\date{Accepted XXX. Received YYY; in original form ZZZ}

\pubyear{2023}

\begin{document}
\label{firstpage}
\pagerange{\pageref{firstpage}--\pageref{lastpage}}
\maketitle

\begin{abstract}
Dark matter may be composed of ultra-light bosons whose de Broglie wavelength in galaxies is $\lambda\sim 1\,{\rm kpc}$. The standard model for this fuzzy dark matter (FDM) is a complex scalar field that obeys the Schr\"{o}dinger-Poisson equations. The wavelike nature of FDM leads to fluctuations in the gravitational field that can pump energy into the stellar components of a galaxy. Heuristic arguments and theoretical analyses suggest that these fluctuations can be modelled by replacing FDM with a system of quasiparticles (QPs). We test this hypothesis by comparing self-consistent simulations of a Schr\"{o}dinger field with those using a system of QPs in one spatial dimension. Simulations of pure FDM systems allow us to derive a phenomenological relation between the number of QPs that is required to model FDM with a given de Broglie wavelength. We also simulate systems of FDM and stars and find that the FDM pumps energy into the stars whether it is described by QPs or a Schr\"{o}dinger field with the FDM adiabatically contracting and the stellar system adiabatically expanding. However, we find that QPs overestimate dynamical heating.

\end{abstract}



\section{Introduction}\label{intro}

In the fuzzy dark matter (FDM) scenario, dark matter is a scalar field $\psi$ whose mass $m$ is so small that its de Broglie wavelength $\lambda = h/mv$ is on the order of a kiloparsec. It therefore exhibits wavelike behavior on galactic scales \citet{hu2000}, which could help distinguish it from ordinary cold dark matter (CDM). Indeed, it was proposed to solve purported problems with CDM such as the apparent deficit of satellites in Milky Way-sized galaxies. Though these problems may actually reflect gaps in our understanding of baryonic physics (see, for example, \citet{bullock2017}) the level of interest in FDM remains high in part because FDM dynamics can lead to phenomena that allow one to distinguish it from other dark matter candidates (see \citet{Hui_2017, neimeyer2020, Hui_2021} for recent reviews). In particular, the interference patterns in an FDM field result in rapid, large-scale fluctuations in the gravitational potential and hence the exchange of energy between the stellar and dark matter components of a galaxy \citep{schive2014,dutta2021,yavetz2022,dutta2023}.

Typically, FDM is treated as a non-relativistic, complex field $\psi$ that obeys the Schr\"odinger equation with a potential given by the Newtonian gravitational potential $\Phi$. Furthermore, FDM contributes $\rho_\psi \propto |\psi|^2$ to the density and thus acts as a source for the gravitational potential through Poisson's equation. We refer to this model of FDM as a Schr\"{o}dinger field (hereafter SF). In the limit where the de Broglie wavelength is much smaller than the scales of interest, FDM behaves like ordinary collisionless dark matter. In fact, one can define an effective phase space distribution function that obeys the collisionless Boltzmann equation in the limit where the de Broglie wavelength goes to zero (see \citet{Widrow_1993} and Section 2.3 below).

Much of our intuition regarding the dynamics of stars in FDM-dominated galaxies comes from the hypothesis that FDM behaves like a system of quasiparticles (QPs). The transfer of energy between FDM and stars can therefore be thought of as arising from two-body interactions between the light stars and heavy QPs. Dimensional analysis suggests that the mass $m_{\rm QP}$ of a QP is given by the de Broglie volume $\sim \lambda^3$ times the local dark matter density \citep{Hui_2017, Church_2019, Hui_2021}. Alternately, one can write $m_{\rm QP}\simeq \left (\hbar/m\right )^3 f({\bf x},{\bf v})$ where $f$ is the phase space density or distribution function (DF) for the dark matter and $\left (\hbar/m\right )^3$ is the phase space volume of a QP. In other words, the number density of QPs in phase space is constant while the mass of each QP is determined by the local DF. If $r_t$ and $v_t$ characterize the extent of the dark matter distribution in space and velocity, then the phase space volume occupied by the dark matter will be ${\cal O}\left (r_t^3 v_t^3\right )$ and the number of QPs will be $N_{QP} \simeq \left (mr_t v_t/\hbar\right )^3$. Theoretical arguments based on the QP hypothesis have been used to estimate the time scale for FDM to disrupt star clusters and wide binary stars, thicken stellar discs, and draw energy out of the orbits of supermassive black holes (see \citet{Hui_2017} and references therein). \citet{amorisco2018} implemented a version of the QP hypothesis in their numerical simulations of the disruption of stellar streams by FDM.

Ultrafaint dwarf galaxies are a particularly attractive arena for studying the effects of FDM on stellar dynamics because they are dominated by dark matter and because the dynamical times are very short. Indeed, their very existence may rule out large regions of FDM parameter space. \citet{Dalal_2022} reach this conclusion using both a heuristic argument and N-body simulations. For the former, they assume that the density fluctuations in a SF are of order unity on the scale of the de Broglie wavelength and show that the change in the variance $\Delta\sigma_*$ of stars over time $t$ in an FDM halo can be written
\begin{equation}
    \Delta\sigma_*^2 \simeq 9\left (\frac{\sigma_*}{\sigma_{dm}}\right )^4\left (\frac{\hbar}{m}\right )^3 \frac{t}{r_{1/2}^4}
    \label{eq:deltasigma}
\end{equation}
where $\sigma_{dm}$ is the velocity dispersion of the dark matter and $r_{1/2}$ is the stellar half-mass radius. \citet{Dalal_2022} use equation \ref{eq:deltasigma} to estimate the time it would take for $\sigma_\star^2$ to double in the UFDs Segue 1 \citep{belokurov2007} and Segue 2 \citep{belokurov2009} as a function of $m$. In doing so, they estimate the sensitivity of stars in these two UFDs to heating by FDM. Of course, both $\sigma_\star$ and $r_{1/2}$ are also changing and therefore equation \ref{eq:deltasigma} should be properly read as a contribution to the Fokker-Planck equation for the stellar DF. If changes in the stellar system are adiabatic, then the virial relation between $\sigma_\star$, $r_{1/2}$ and $M$ will be approximately satisfied at all times. Whether $\sigma_\star$ increases, decreases, or stays the same will therefore depend on the radial density profile of the system and whether it changes with time. 

\citet{Dalal_2022} obtained more quantitative results by simulating FDM-star galaxies. However, they treated the gravitational potential as fixed so that the FDM field could be written as a superposition of energy eigenstates, which evolved independently via a unitary, analytic transformation. This greatly reduced the computational complexity of their simulations though they forfeited self-consistency in computing the gravitational potential as well as energy conservation.

Recently, \citet{dutta2023} conducted self-consistent simulations of stellar systems that were embedded in FDM halos. The FDM halos were extracted from the cosmological simulation described in \citet{schive2014} while the stars were introduced by treating them as a test-particle sample drawn from an equilibrium distribution function, which in turn was constructed using the time and azimuthally-averaged FDM density profile. Since the stars in their simulations accounted for less than $0.1\%$ of the total mass, this procedure yielded a system in approximate dynamical equilibrium. \citet{dutta2023} found that the stars were rapidly heated by the fluctuating FDM potential. In particular, over $7.5\,{\rm Gyr}$ the characteristic size of the stellar components increased by an order of magnitude while the velocity dispersion increased by a factor of $2-3$. The results therefore confirmed the arguments in \citet{Dalal_2022} though they cautioned that their simulations made a number of assumptions that might not apply to actual UFDs.

\citet{El_Zant_2019, Bar_Or_2019} and \citet{Bar_Or_2021} provided more formal treatments of stellar heating by FDM within the Fokker-Planck formalism. These studies confirmed the correspondence between heating of stars by QPs and by a SF. However, these analyses were carried out for homogeneous systems and were therefore not directly applicable to isolated systems of FDM and stars.

The goal of this paper is to test the QP hypothesis for isolated, self-gravitating systems by running side-by-side SF and QP simulations. We consider pure SF and pure QP systems as well as systems that have both dark matter and stars. All components are live and the gravitational field is determined self-consistently. All of our simulations are done in one spatial dimension and assume plane symmetry. This choice is made to keep the computational complexity at a manageable level since the complexity in simulating a self-gravitating SF in three dimensions is very high when the de Broglie wavelength is small. Furthermore, one can easily visualize the full position-velocity phase space in one dimension. 

It is worth reflecting on the applicability of our results to three dimensions. Self-gravitating systems in one dimension share many properties with their three dimensional counterparts. In both cases, equilibrium systems can be set up via the Jeans theorem as described in \citet{BT2008} and below. We choose the lowered isothermal plane as initial conditions \citep{weinberg1991}, which is the one dimensional analog of the well-known King model \citep{king1966}. We further note that both one and three dimensional systems exhibit vibrations about these equilibrium states (see \citet{kalnajs1973, mathur1990, weinberg1991, Widrow_2015} for the one dimensional case and \citet{BT2008} and references therein for the three dimensional case). On the other hand, systems that start from cold initial conditions will undergo gravitational collapse and phase mixing regardless of their dimensionality. (See \citet{fillmore1984} for a nice example in the context of cosmological structure formation.) Of course, the force-law in a plane-symmetric one-dimensional system is very different from the three dimensional force-law. In particular, the force between two infinite planes (the one-dimensional analog of a three-dimensional particle) is independent of the separation between the particles. The scattering of two planes therefore occurs instantaneously when they pass one another and the force changes sign. Thus the details of two-body relaxation will be rather different in one and three dimensions though the essential physics of the process, which involves the transfer of energy between heavy particles and a sea of light particles, is the same.

At first glance, it is not obvious whether standard N-body methods are appropriate for QPs. In particular, the density fluctuations associated with FDM appear and disappear in a stochastic manner. Can these fluctuations really be described by particles that follow continuous Newtonian orbits? One might imagine a scheme in which QPs are randomly created and destroyed but adding such complications would seem to defeat the purpose of using QPs to model FDM.

A further examination of equation \ref{eq:deltasigma} suggests that a simple N-body scheme may actually capture the dynamics of QPs. We define the crossing time of the system as $t_{cr}\equiv r_{1/2}/\sigma_*$. The fractional change in the stellar variance over a single crossing time is then
\begin{equation}
\frac{\Delta\sigma_*^2}{\sigma_*^2} \simeq 9 \frac{\sigma_*}{\sigma_{dm}}
\left (\frac{\hbar}{mr_{1/2}\sigma_{dm}}\right )^3.
\label{eq:fractionalsigma}
\end{equation}
The quantity $(mr_{1/2}\sigma_{dm})^3$ is, up to a constant of order unity, equal to the phase space volume of the system while $\hbar^3$ is the volume of a quantum cell in phase space. Thus, the final factor in equation \ref{eq:fractionalsigma} is equal to $1/N_{\rm cell}$, where $N_{\rm cell}$ is equal to the number of cells in the system, again up to a constant of order unity. This expression is reminiscent of the usual two-body relaxation formula if we identify $N_{\rm cell}$ with the number of ``particles" in the system \citep{BT2008} and motivates the idea that QPs have equal phase space volume. The mass of each QP is then given by the local value of the DF multiplied by this volume.

As we'll see, the QP hypothesis works extremely well for systems that stay close to their equilibrium distribution. For example, the spectrum of vibrations about an equilibrium state is remarkably similar whether the system is modelled as a SF or system of QPs. Conversely, when the system evolves significantly away from its initial state, the behaviour of a SF and a system of QPs can be very different. In particular, we find that over long times, QPs are more efficient at heating stars than a SF, a result that may have implications for understanding the evolution of dwarf galaxies in an FDM cosmology.

In Section 2, we describe various numerical methods used in our simulations. In Section 3, we present results for a pure FDM system modelled either as a SF or QPs. In both cases, we find that the system vibrates due to fluctuations in the density and potential. The amplitude of the vibrations can be characterized by fluctuations in the virial ratio ${\cal V} = 2K/W$ where $K$ and $W$ are the kinetic and potential energies of the system. We find that the fluctuations in ${\cal V}$ are approximately proportional to $\lambda^{1/2}$ or, alternatively, $N_{QP}^{-1/2}$, as one would expect from root-N statistics. These results allow us to derive a correspondence between $N_{QP}$ and $m$. We also examine the power spectra for the gravitational force in each of the simulations. Results from simulations that include stars and either FDM or QPs are given in Section 4. We show that in both cases, the stars gain energy at the expense of the FDM or QPs. Nevertheless, the stars maintain approximate virial equilibrium throughout with no long term drift in ${\cal V}$. We conclude in Section 5 with a summary of our results and some thoughts on their applicability to systems in three dimensions.

\section{Preliminaries}

In this section we present the numerical methods used to set up and run simulations with a SF, a system of QPs, and stars. In all of our simulations we start from equilibrium initial conditions. To this end, we first write down a DF that solves the time-independent collisionless Boltzmann equation. For the SF, we set up initial conditions for the wave function $\psi$ so that its phase space representation is approximately equal to the same equilibrium DF used for the QPs and stars. 

\subsection{Distribution Function}

Consider a system in one dimension with DF $f=f(x,v)$. In a static potential $\Phi(x)$ any function of the specific energy (i.e., energy per unit mass) $E = \frac12 v^2 + \Phi$ will be a solution to the time-independent collisionless Boltzmann equation. In this work, we take the initial DF to be that of the lowered isothermal plane \citep{weinberg1991}. This model is a truncated version of the isothermal plane first considered by \citet{spitzer1942} and \citet{camm1950} and has been used to describe the vertical structure of disc galaxies such as the Milky Way. In the original model, the phase space density went to zero only in the limit $|x|\to \infty$ or $|v|\to \infty$. In the lowered isothermal plane, the phase space density goes to zero at finite $x$ and $v$ and is therefore suitable as a starting point for numerical simulations. The DF for the lowered isothermal plane is given by 
\be
f(x,v) = f(E) = 
\begin{cases}
f_0\left (e^{-E/\sigma^2} - e^{-E_0/\sigma^2}\right ) & E<E_0\\
 0 & E\ge E_0~.
\end{cases}
\label{eqn: DF}
\ee 
where $f_0$ is a normalization constant, $\sigma$ is a velocity scale, and $E_0$ is the energy cut-off. Note that $f$ has physical dimensions of $[{\mathsf M}][\mathsf L]^{-3}[{\mathsf L/T}]^{-1}$. If we integrate the DF over the $x-v$ volume, we obtain the total surface density $\Sigma$, which is the analog of the total mass in a three-dimensional system. The density $\rho$ is found by integrating $f$ over $v$
\begin{align}
\rho(\Phi) & = 
    2\int_0^{v_m} f(E) dv \nonumber\\
    & = \sqrt{2\pi}\sigma f_0\left [{\rm erf}\left (u_m\right )e^{-\Phi/\sigma^2} - \frac{2}{\sqrt{\pi}} u_m e^{{-E_0/\sigma^2}}\right ]
\end{align}
where $v_m(x)\equiv \sqrt{2\left (E_0-\Phi(x)\right )}$ is the maximum velocity at position $x$ and $u_m\equiv v_m/\sqrt{2}\sigma$. 
Likewise, the variance of the velocity as a function of the potential is
\begin{align}
\langle v^2\rangle & =  
    \frac{2}{\rho}\int_0^{v_m} v^2 f(E) dv \nonumber\\
    & = \sigma^2
    \frac{{\rm erf}(u_m) - \frac{2}{\sqrt{\pi}}e^{-u_m^2}
    \left (u_m + \frac{2u_m^3}{3}\right )}{{\rm erf}(u_m) - 
    \frac{2}{\sqrt{\pi}}e^{-u_m^2}u_m }.
\end{align}
The density, force, and potential are found as functions of $x$ by numerically integrating Poisson's equation. We define $x_t$ as the position at which the density goes to zero. The total surface density is then given by $\Sigma = 2\int_0^{x_t} dx\rho(x)$. We choose units in which $G= \sigma = 1$ and adjust $f_0$ so that $\Sigma = 1/\pi$. We then define the unit of length to be $x_0 \equiv \sigma^2/\pi G\Sigma=1$. The velocity dispersion is maximal at $x=0$ with a value that depends on $\sigma$ and $E_0$ and decreases smoothly to zero as $x$ approaches $x_t$.

In the limit $E_0\to \infty$, the model reduces to the isothermal plane \citep{spitzer1942, camm1950}. In that limit the potential and density are elementary functions of $x$ and the velocity dispersion is a constant and equal to $\sigma$. For example, the potential for the isothermal plane is given by
\be\label{eq:potential}
\Phi(x) = 2\,{\rm ln}(\cosh{x})~~~~~~~~(E_0\to \infty).
\ee
In what follows, we use $E_0=3$. We then find that $x_t\simeq 2.00$ and $v_t \equiv\sqrt{2E_0}\simeq 2.45$. For $x=0$, the density and velocity dispersion are $\rho_0 \equiv \rho(x=0) =  0.204$ and $\langle v^2\rangle^{1/2}(x=0) = 0.78$. The time for a particle of energy $E$ to complete one orbit in the $x-v$ plane is
\begin{equation}
T(E) = 4\int_0^{x_t} \frac{dx}{\sqrt{2(E-\Phi(x)}},
\end{equation}
which is a monotonically increasing function of $E$.  We define the dynamical time as $t_{\rm dyn}\equiv T(E\to 0)$. Since $\Phi(x)\simeq 2\pi\rho_0 x^2$ for $x\ll 1$, we have $t_{\rm dyn}=\sqrt{\pi/\rho_0}\simeq 3.92$. Finally, the system occupies a phase space volume
\begin{equation}
    \Omega = 4\int_0^{x_t} dx \sqrt{2(E-\Phi(x)},
\end{equation}
which is just the usual action divided by $m$. Thus, if there are $N_{QP}$ QPs, each particle will occupy a phase space volume $\Omega_{QP} = \Omega/N_{QP}$. For $E_0=3$, we find $\Omega\simeq 14.5$.

\subsection{Equations of Motion}

In the SF model, FDM obeys the Schr\"{o}dinger-Poisson system of equations. In one dimension we have
\be
i\frac{\partial \psi}{\partial t} = -\frac{\mathcal{R}}{2}\frac{\partial^2 \psi}{\partial x^2} + \frac{1}{\mathcal{R}}\Phi\psi
\label{eqn: Schrodinger}
\ee
where we have introduced the dimensionless parameter ${\cal R} \equiv \hbar/m\sigma x_0 = \hbar/m$. The QPs and stars are modelled as point particles whose orbits are determined by Newton's second law. 

The density associated with $\psi$ is given by $m|\psi|^2$. For stars and QPs, we calculate the density $\rho_p$ using a particle mesh scheme where the mass of each particle is assigned to the cell whose center is closest to the position of the particle. We then have
\be
\frac{\partial^2\Phi}{\partial x^2} = 4\pi (m |\psi|^2 + \rho_p).
\label{eqn: Poisson}
\ee

Numerical evolution of FDM is handled by the kick-drift-kick scheme that is now widely used in both cosmological simulations and simulations of isolated systems \citep{woo2009,Edwards_2018}. The Poisson equation is solved using standard Greens function and FFT methods with zero-padding to handle the boundary conditions of an isolated system \citep{hockney2021computer}.

\subsection{Initial Conditions}

Our goal is to set up an initial state $\psi(x)$ whose phase space representation is approximately equal to equation \ref{eqn: DF}. We do so using the algorithm outlined in \citet{Widrow_1993}. To make the connection between $\psi$ and $f$ we use the Husimi Q transformation \citep{husimi1940} in which the effective DF is the absolute square of the windowed Fourier transform of $\psi$:
\begin{equation}
F(x,v) = |\Psi(x,v)|^2
\end{equation}
where
\begin{equation}\label{eq:husimi}
\Psi(x,v) = {\cal N} \int e^{-(x-x')/2\eta^2} e^{ix'v/{\cal R}}\psi(x')dx'
\end{equation}
and ${\cal N}$ is a normalization constant. In the limit $\hbar\to 0$, $F$ obeys the familiar continuity and Jeans equations for a system of collisionless particles \citep{Madelung1926,skodje1989,Widrow_1993}. In general we choose $\eta = \left ({\cal R}x_t/v_t\right )$, which gives equal resolution in space and velocity.

To evaluate equation \ref{eq:husimi}, we set up a phase space grid with 
\begin{equation}
x_n = \Delta_x n - x_{\rm max}~~~~~~~n=0\dots N-1
\end{equation}
and 
\begin{equation}
v_k = \Delta_v k - v_{\rm max}~~~~~~~k=0\dots N-1
\end{equation}
where $\Delta_x \equiv 2x_{\rm max}/N$ and $\Delta_v\equiv 2v_{\rm max}/N$. We set $x_{\rm max}$ and $v_{\rm max}$ to be 1.5 times larger than $x_t$ and $v_t$ so that the grid can also be used to map $\psi$ in phase space even after the system has evolved away from its initial state. Equation \ref{eq:husimi} then takes the form of a discrete Fourier transform provided we set $N = 2x_{\rm max}v_{\rm max}/\pi {\cal R}$.

We next consider the ansatz
\begin{equation}
    \psi(x_n) = \frac{1}{\sqrt{N m}}\sum_{k} \sqrt{f(x_n,v_k)} R_k e^{i x_n v_k/{\cal R}} ~,
    \label{eqn: WidrowKaiser Trick}
\end{equation}
where $\{R_k\}_k$ is a sample of a random variable uniformly distributed on the complex unit-circle. A straightforward calculation shows that the phase space representation for $\psi$ yields $F\simeq f$ \citep{Widrow_1993}.

As discussed in the introduction and also \citet{El_Zant_2019,Bar_Or_2019} and \citet{Bar_Or_2021}, in three dimensions, we expect each QP to occupy a (position-velocity) phase space volume of $\left (\hbar/m\right )^3$. In one dimension, we therefore expect QPs to occupy an $x-v$ phase space volume of $\hbar/m$. The surface density of each QP is then given by $\hbar/m$ times the distribution function $f$. To initialize the QP distribution we uniformly sample the phase space volume $\Omega$ for points $\{(x_j,p_j)\}_{j}$. The $j^\text{th}$ QP resides at the coordinates $(x_j,p_j)$ and is given surface density ${\cal N}f(x_j,p_j)$ where the normalization factor ${\cal N}$ is adjusted so that the total QP surface density is equal to $\Sigma$. The QP distribution is then rather different from the normal distribution in N-body simulations where simulation particles, at least for a particular system such as a disc, bulge, or satellite, have the same mass. In that case, regions of high phase space density have a higher density of simulation particles. In the QP case, phase space is uniformly sampled but the particles have different masses, or rather different surface densities since we are in one dimension. In this regard, our QP simulations are similar to multi-mass N-body simulations, as discussed, for example, in \citet{sigurdsson1995}. For collisionless simulations, multi-mass methods afford one higher mass resolution for fixed computational cost. In our case, it's the collisional properties of the QPs that we wish to study while the scheme chosen is motivated by the physics of FDM.

\section{SF versus QPs in FDM simulations}\label{sec: FDM v QPs Pure}

In this section we compare the evolution of SF and QP systems by running two sequences of simulations with different values of either ${\cal R}$ or $N_{QP}$. In all of the simulations, the initial conditions are derived from equation \ref{eqn: DF} with $E_0=3$, $\sigma=1$, and $f_0$ adjusted so that $\Sigma=1/\pi$. The simulations are run up to $T = 96\approx 24 t_{\rm dyn}$. We solve Poisson's equation on a mesh using a grid with $N$ cells. We set $N=500$ for QP simulations. As discussed in the introduction, the number of cells for the SF simulations is determined by the ratio of $\lambda\sim \hbar/m\sigma_{dm}$ to the size of the system ($\sim r_{1/2}$). The timestep $h$ is chosen to satisfy the Courant condition $h = 0.5 \Delta x / v_t$. Note that while this gives $h\approx0.0025$ for QPs, in the case of the SF the mesh-size changes between simulations, and thus different values of $h$ are used. This is a choice that allows us to lower the computational time of SF simulations at higher values of $\cal R$.

\subsection{Oscillations}
\begin{figure}
    \centering
    \includegraphics[width = \columnwidth]{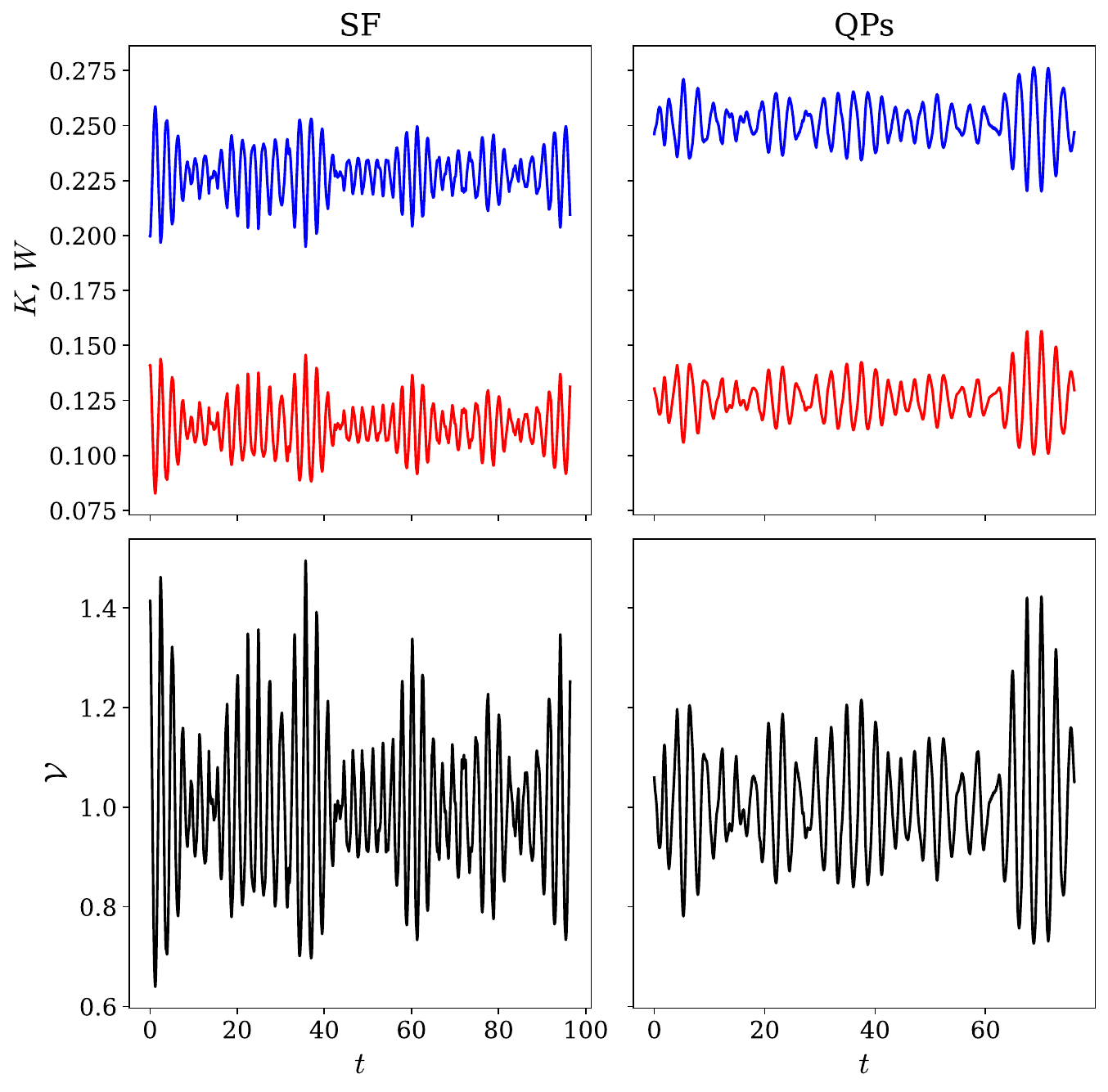}
    \caption{Kinetic and potential energies and the virial ratio as a function of time for the SF (left column) and QPs (right column). The upper panels show $K$ (red) and $W$ (blue) while the lower panels show ${\cal V} \equiv 2K/W$.  }
    \label{fig: Section3 Energy Curves}
\end{figure}

Both SF and QP systems oscillate about their equilibrium configurations. To illustrate this, we choose two independent QP and SF systems with parameters $N_{QP}=100$, $h \approx 0.0025$, $N = 500$ and ${\cal R} \approx 0.067$, $h \approx 0.0148$, $N=84$ respectively. In Fig. \ref{fig: Section3 Energy Curves}, we plot the time evolution of the kinetic and potential energies, $K$ and $W$, as well as the virial ratio ${\cal V}\equiv 2K/W$. The oscillations in $K$, $W$, and ${\cal V}$ are qualitatively very similar in the two systems. The dominant period is approximately $P \simeq 3 \simeq 0.8t_{\rm dyn}$ while the beat patterns indicate that there are multiple oscillations with different frequencies. These oscillations are characteristic of one-dimensional self-gravitating systems and have been studied by \citet{kalnajs1973, mathur1990, weinberg1991} and \citet{Widrow_2015}.

\begin{figure}
    \centering
    \includegraphics[width = \columnwidth]{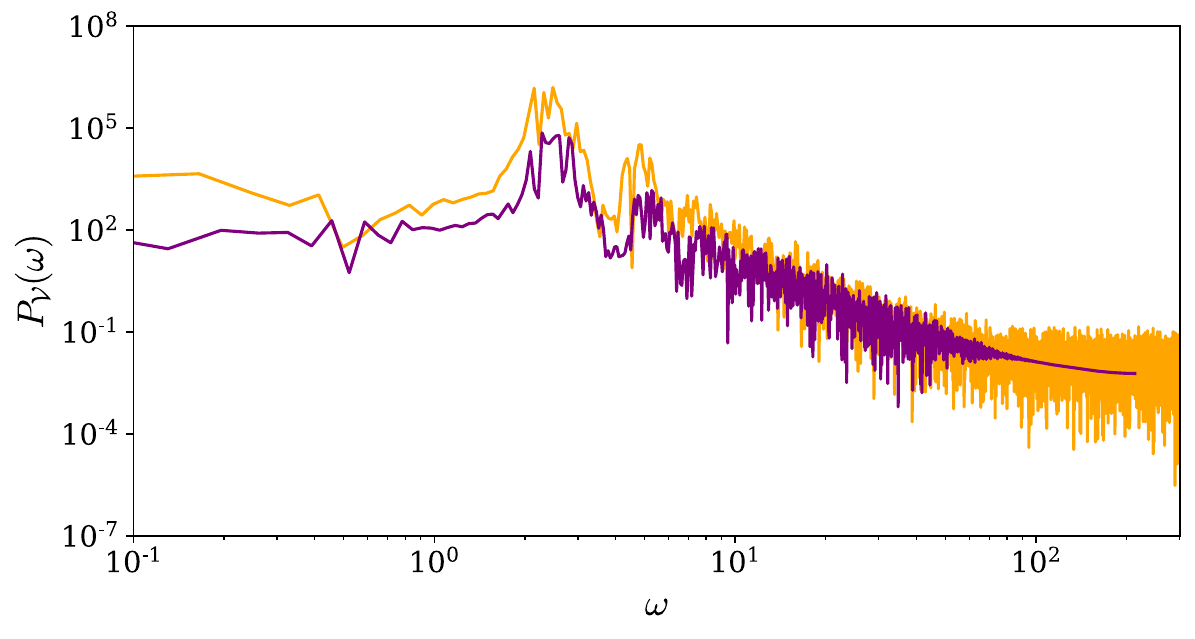}
    \caption{Temporal power spectra of ${\cal V}$ for the SF (purple) and QPs (orange).}
    \label{fig:virialpowerspectrum}
\end{figure}

To further illustrate the similarities between SF and QP dynamics, we show the temporal power spectra $P_{\cal V}(\omega)$ of ${\cal V}$ in Fig. \ref{fig:virialpowerspectrum}. The spectra are remarkably similar. In particular, there is a prominent peak at $\omega\simeq 2$, which corresponds to a period of $P \simeq 3$, as expected from Fig.\ref{fig: Section3 Energy Curves}. In addition, there are secondary peaks at multiples of the first one as well as a continuous distribution with $P\propto \omega^{-4}$ at large $\omega$. We'll return to this last point below.

\begin{figure}
    \centering
    \includegraphics[width = \columnwidth]{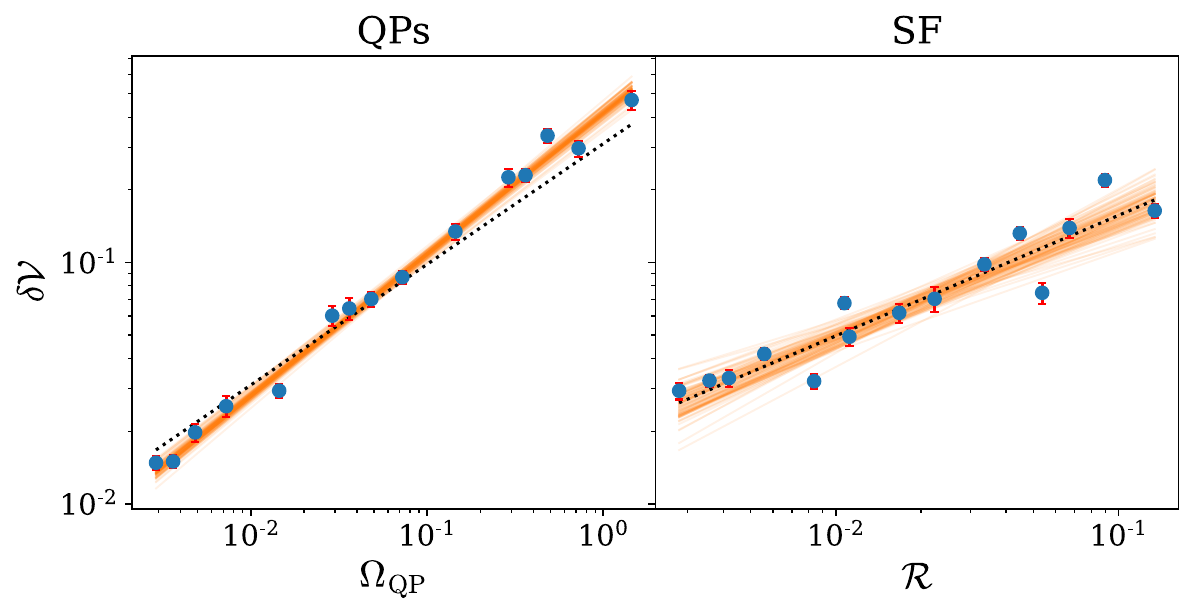}
    \caption{Curve fits of the RMS amplitude's of oscillation in $\cal V$, plotted against either $\Omega_{\mathrm{QP}}$ or $\mathcal{R}$ on log-log axes. The dotted black curve is the mean fit with the power ($p$ or $q$) fixed to $1/2$. The orange lines correspond to a sample of $p$ and $q$ values from the emcee routine. Each data point is the average of 15 trials for the given parameter. Similarly, the uncertainties of each data point are the standard deviation of each sample of trials.}
    \label{fig: RMS amplitude fits}
\end{figure}

To derive a phenomenological relation between ${\cal R}$ and $\Omega_{QP}$ we run an SF sequence with $0.003 < {\cal R} < 0.15$ and a QP sequence with $0.003 < \Omega_{QP} < 1.5$. In the case of the latter, the number of QPs varies between $10$ and $5000$. In each simulation, we calculate  $\delta {\cal V} \equiv \langle \left ({\cal V}-1\right )^2\rangle^{1/2}$, that is, the root mean square deviation in ${\cal V}$ from unity over the course of the simulation. For each value of ${\cal R}$ or $\Omega_{QP}$, we run 15 simulations with different realizations of the initial conditions and compute the mean and standard deviation of $\delta{\cal V}$. We then model $\delta {\cal V}$ as a power-law function of ${\cal R}$ or $\Omega_{QP}$:
\begin{equation}
    \delta{\cal V}_{FDM} = b{\cal R}^q
\end{equation}
and
\begin{equation}
    \delta{\cal V}_{QP} = a{\Omega}_{QP}^p.
\end{equation}
The parameters $a,p,b$, and $q$ are determined by fitting $\log{\cal V}$ to a straight line using the Markov chain Monte Carlo sampler \textrm{emcee} \citet{emcee2013}. The results are shown in Fig. \ref{fig: RMS amplitude fits} and the fit parameters are given in Table 1. 

Since the fluctuations are seeded by Poisson noise from the initial conditions, we anticipate that $\delta{\cal V}$ will be proportional to ${\cal R}^{1/2}$ or $\Omega_{\mathrm{QP}}^{1/2}$. This expectation is borne out in the SF simulation. In the case of the QPs, we find $p\simeq 0.58\pm 0.015$. Though the discrepancy from $p=0.5$ is statistically significant, it amounts to only a $\pm 15\%$ change in $\delta{\cal V}$ over two orders of magnitude in $\Omega_{\mathrm{QP}}$. Table \ref{tab: Parameters} also includes values for $a$ and $b$ when $p$ and $q$ are fixed to $0.5$

\begin{figure}
    \centering
    \includegraphics[width = \columnwidth]{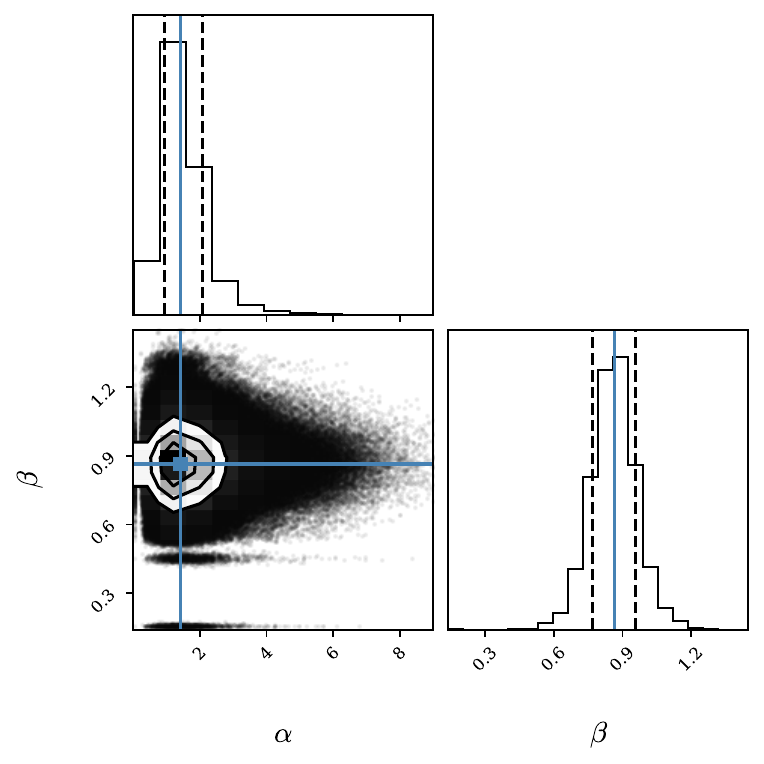}
    \caption{Corner plot showing the joint and marginal distributions of $\alpha$ and $\beta$. The blue lines represent the 50th percentile, while the dashed black lines are the 0.16 and 0.84 quantiles, representing $\pm 1 \sigma$.}
    \label{fig:cornerplot}
\end{figure}

Setting $\delta{\cal V}_{\mathrm{FDM}}$ equal to $\delta{\cal V}_{\mathrm{QP}}$ leads to the power-law relation 
\begin{equation}\label{eq:NscriptR}
\Omega_{\mathrm{QP}} = \frac{\Omega}{N_{QP}} = \alpha {\cal R}^{\beta}~.
\end{equation}
The joint and marginal probability distribution functions for $\alpha$ and $\beta$ are shown in Fig. \ref{fig:cornerplot} while best-fit values for $\alpha$ and $\beta$ are given in Table 1. 

\begin{table}
    \centering
    \renewcommand{\arraystretch}{1.4}
    \begin{tabular}{l|l|l} 
  Parameters  & Values & $p=q \equiv 1/2$\\ \hline \hline
$a$ & $0.412^{+0.021}_{-0.021}$ & $0.311^{+0.017}_{-0.017}$ \\ 
$p$ & $0.583^{+0.015}_{-0.015}$ & $0.5$ \\
$b$ & $0.499^{+0.123}_{-0.102}$ & $0.496^{+0.030}_{-0.030}$\\
$q$ &  $0.502^{+0.054}_{-0.052}$ & $0.5$\\ \hline
$\alpha$ & $1.409^{+0.669}_{-0.483}$ & $2.555^{+0.452}_{-0.383}$ \\
$\beta$ & $0.864^{+0.094}_{-0.095}$ & 1
    \end{tabular}
    \label{tab: Parameters}
    \caption{Fit parameters for Figure \ref{fig: RMS amplitude fits} and for the relation between $\Omega_{\mathrm QP}$ and ${\cal R}$. The third column shows the values given $p$ and $q$ fixed at $1/2$.} 
\end{table}

\subsection{Force Power Spectra}
\begin{figure}
    \centering
    \includegraphics[width = \columnwidth]{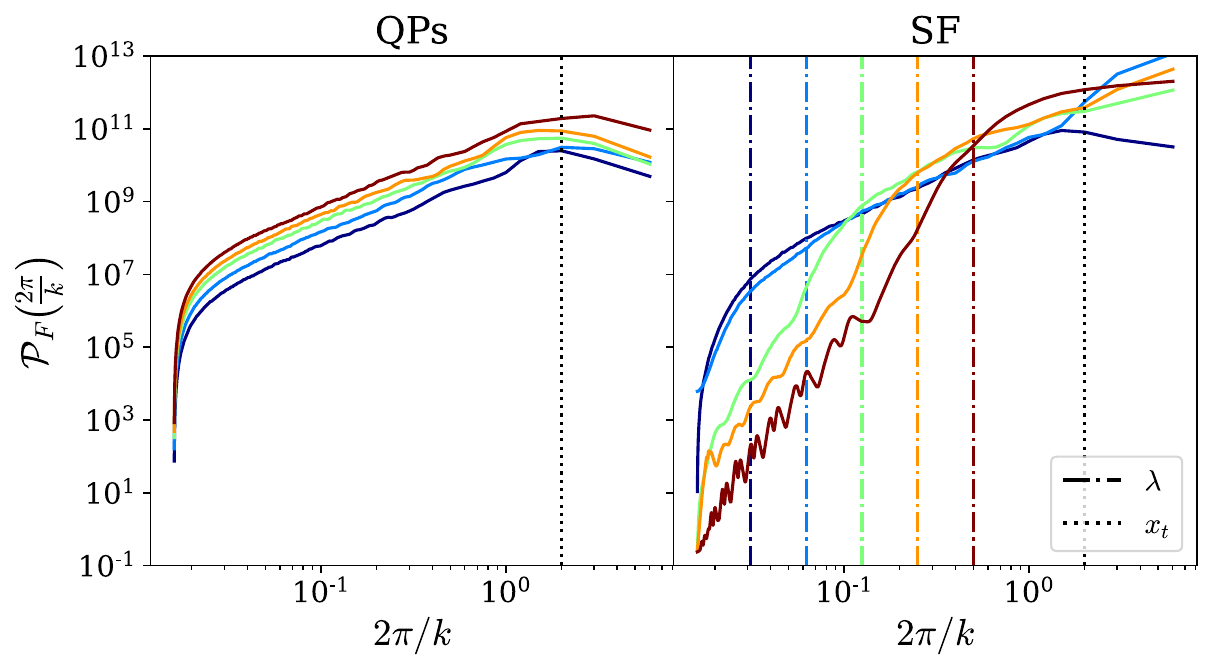}
    \caption{Spatial power spectra of oscillations from the time-independent acceleration field $-\nabla\Phi$, for SFs with differing values of fuzziness ${\cal R}$, and the corresponding number of quasiparticles $N_{\rm QP}({\cal R})$. The colours (dark blue, light blue, green, orange, red) correspond to the cited values of $\cal R$, in increasing order. The dotted black line marks the truncation length $x_t$, while the dashed vertical lines mark the respective de Broglie wavelengths $\lambda$. Note that the horizontal axes represent the wavelength $2\pi/k$, where $k$ is the spatial frequency.} 
    \label{fig: Force Power Spectra}
\end{figure}

In Fig. \ref{fig: Force Power Spectra} we show power spectra for force fluctuations, ${\cal P}_F(k)$, from SF simulations with ${\cal R} \approx 0.0084,0.017,0.034,0.067,0.13$ and from QP simulations with the corresponding values of $N_{\rm QP}$ as given by the relation derived in the previous section. In these simulations, we fix the number of grid points to $N=500$, for a more direct comparison of the power spectra. To calculate ${\cal P}_F(k)$, we first determine the force fluctuations as a function of $x$ and $t$ by subtracting off the time-averaged force:
\begin{equation}
    F'(x,t) \equiv -\frac{d\Phi}{dx} + \left\langle \frac{d\Phi}{dx}\right\rangle_t~.
\end{equation}
We next compute the spatial Fourier transform, which yields $\tilde{F}'(k,t)$. The desired force fluctuation power spectrum is found by taking the time average of $|\tilde{F}'|^2$. Note that ${\cal P}_F$ is just the Fourier transform of the force auto-correlation function, which plays a central role in the Fokker-Planck analyses of \citet{Bar_Or_2019,El_Zant_2019}, and \citet{Bar_Or_2021}.

In the case of QPs, ${\cal P}_F(k)\propto k^{-2}$ down to $k_t \simeq 2\pi / x_t$. The $k^{-2}$ spectrum corresponds to Brownian noise. In one dimension, the gravitational force at position $x_p$ is proportional to the difference between the surface density for $x>x_p$ and the surface density for $x<x_p$. Since the density can be modeled as white noise, its Fourier transform will be constant in $k$ and therefore the Fourier transform for the force or surface density will be proportional to $k^{-1}$. 

Fig. \ref{fig: Force Power Spectra} also provides a connection to the temporal Fourier transform of $\cal V$ found in Fig. \ref{fig:virialpowerspectrum}. Since the collisionless Boltzmann equation is linear and first-order in both space and time derivatives, we expect that $k\propto\omega$ for the two-dimensional (spatial-temporal) Fourier transforms of various quantities such as the potential and force. Furthermore, both the one- and two-dimensional Fourier transform of the force are $-ik$ times the Fourier transform of the potential:
\begin{align}
    \hat{F}(k,\omega) &= -\frac{\partial}{\partial x}\iint \Phi(x,t)e^{-ikx+i\omega t}dxdt \\
    &= -ik \hat{\Phi}(k,\omega) 
\end{align}
Thus, $|\hat{F}(k)|^2=k^2|\hat{\Phi}(k)|^2$ and the $\omega^{-4}$ behaviour for the temporal Fourier transform of ${\cal V}$ is consistent with the $k^{-2}$ behaviour for spatial Fourier transform of $F$.

In our SF simulations the force power spectra exhibit a fall off for length-scales below $\lambda$. Our Fig. \ref{fig: Force Power Spectra} is analogous to results from \citet{Dalal_2021}
(see the $\Delta t=0$ curves in their Figure 13). This damping of power below the de Broglie scale was one of the initial reasons FDM was introduced and meant that in the context of structure formation, FDM had some of the same properties as warm dark matter \citet{hu2000}.

\section{Dynamical heating of stars by FDM}
In this section we compare the dynamical effects of a SF verses QPs on a system of stars. For illustrative purposes, we assume that dark matter and stars each account for half the mass of an isolated system and that both components start with the same equilibrium DF used in the previous section. We run two independent simulations with $5\times 10^4$ particles to represent the stars and either a SF or QPs for FDM. The particles representing the stars all have the same mass; their initial positions and velocities are initialized by sampling the DF using a simple accept-reject algorithm. We set ${\cal R} \simeq 0.067$ for FDM. With this value our previous results suggest using $N_{\rm QP} = 106$ if $p$ and $q$ are taken as free parameters or $N_{\rm QP} = 84$ if $p$ and $q$ are fixed to $1/2$. Here, we set $N_{\rm QP}=106$. In both simulations, we fix the number of grid points to $N=500$, and the time step to $\Delta t \approx 0.0025$.

\subsection{Phase space evolution of stars and dark matter}

\begin{figure*}
    \centering
    \includegraphics[width = \textwidth]{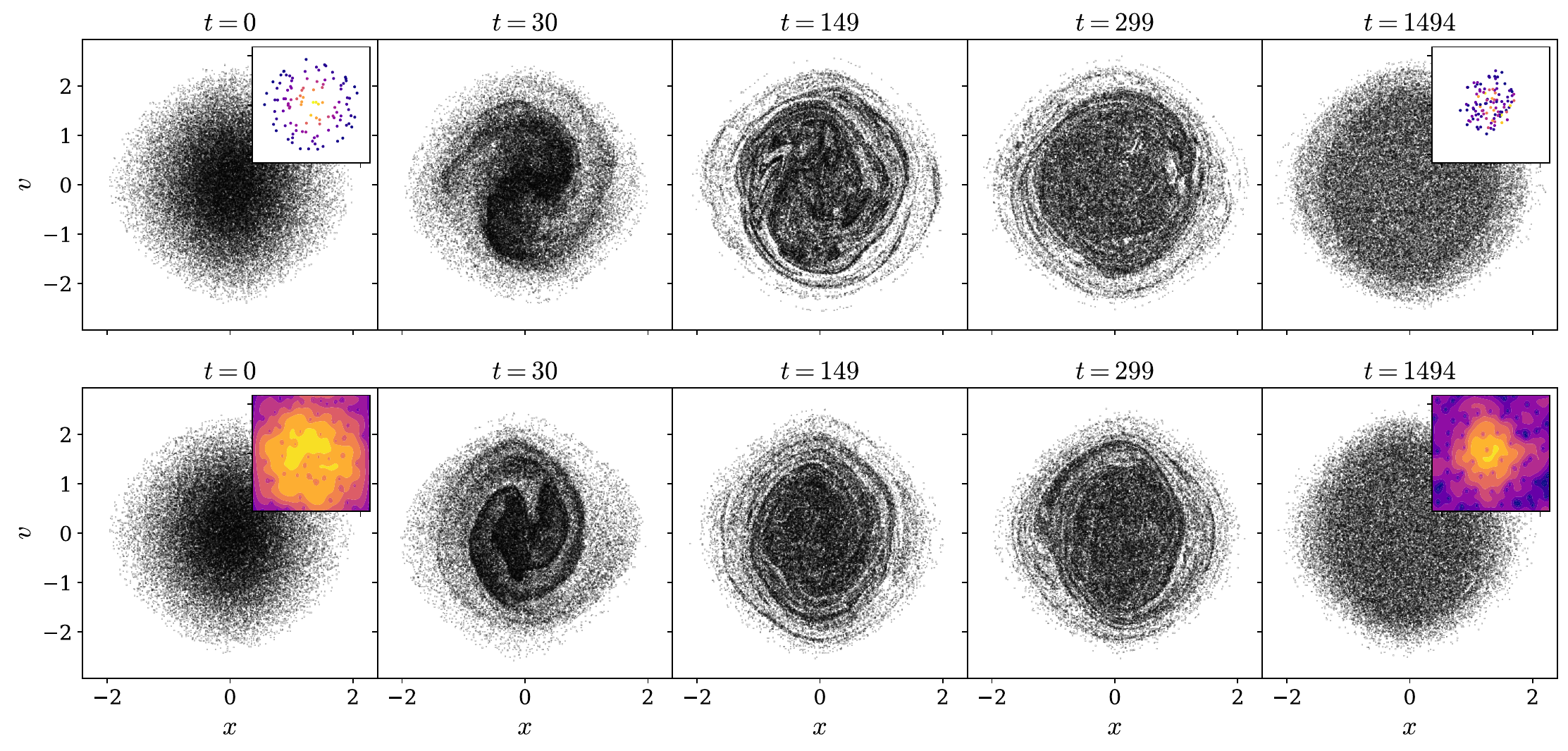}
        \caption{Snapshots of the phasespace distributions of the QPs+stars (top) and SF+stars (bottom), taken at approximately 0, 7.5, 37.5, 75, 375 dynamical times. Particles representing the stars are shown as black points while the QP/SF portions are plotted in the upper-right corners of the first and last snapshots. In the case of QPs, the colour signifies the mass. In the SF case, the colour represents the phase-space density given by the Husimi Transform, plotted as countours at 15 levels on a log-scale.}
    \label{fig: Phasespace Evolution V2}
\end{figure*}

In Fig. \ref{fig: Phasespace Evolution V2} we present a sequence of phase space snapshots for the SF and QP runs. In both cases, the evolution proceeds through three stages: an initial isothermal phase during which the DF is peaked at the origin of the $x-v$ plane; an intermediate phase where the DF is disturbed by either the SF or QPs; a final phase when the DF is approximately constant for $E$ less than the truncation energy. Interestingly enough, the edge of the distribution doesn't change by very much over the course of the simulation. Rather, there appears to be a redistribution of particles from small to large energies. 

\begin{figure}
    \centering
    \includegraphics[width = \linewidth]{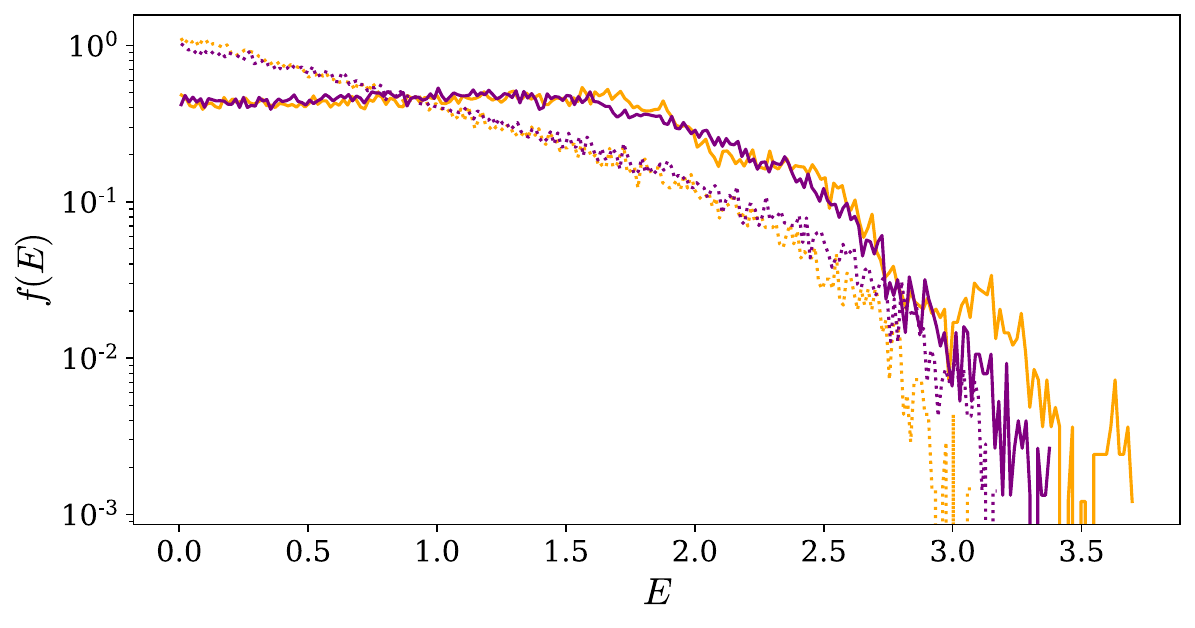}
    \caption{Initial (dotted) and final (solid) distributions of specific energies among particles in either mixed regime; QPs+stars (orange) or SF+stars (purple). These curves exclude the energies of the QP/SF component.}
    \label{fig: E DFs}
\end{figure}

\begin{figure}
    \centering
    \includegraphics[width = \columnwidth]{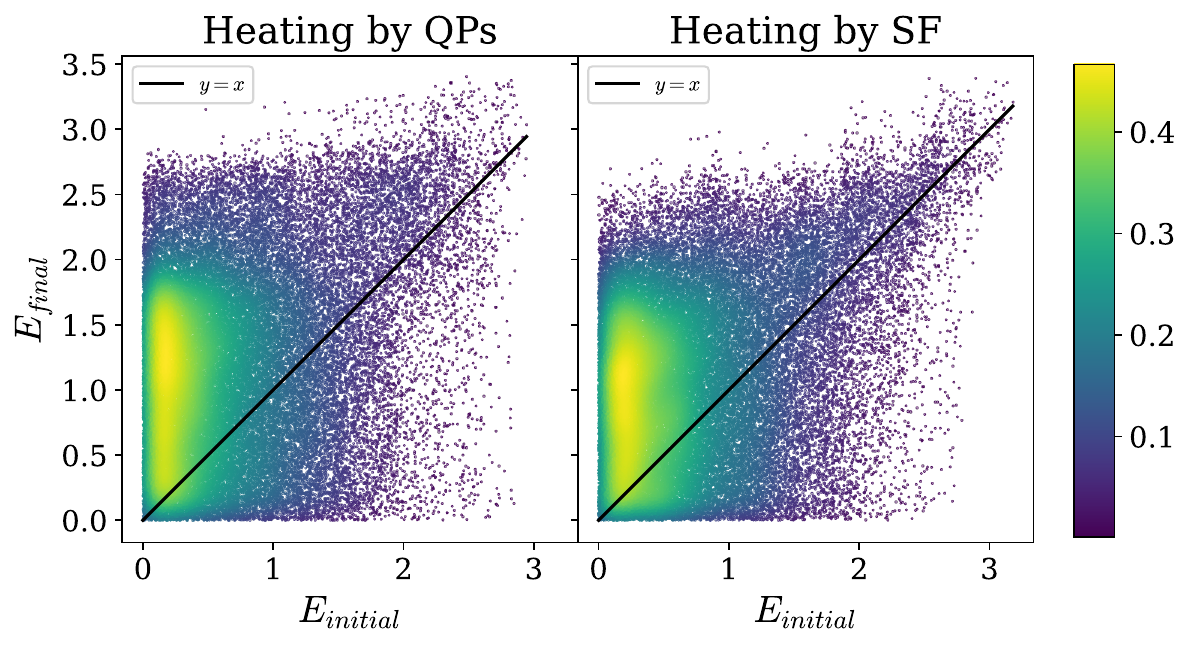}
    \caption{Scatter plot of the initial vs final specific energies of every particle heated by QPs (left) or SF (right). The color scale is linear in density in the $E_i-E_f$ space. (The overall scale of the density is irrelevant.}
    \label{fig: Scatter E}
\end{figure}

These points are further illustrated in Fig. \ref{fig: E DFs}. The energy distribution $f(E)$ shows a transition from an isothermal or Maxwellian distribution at the start of the simulation to one that is approximately constant in energy for $E\lesssim 1.5$ . The shift of particles from low-to-high energy is shown in greater detail by the scatter plots of Fig. \ref{fig: Scatter E}. While the distribution of initial and final energies show an overall increase in energy, it is clear that the heating effect is more pronounced for the particles with $E_{\rm initial}\lesssim 1$.

\begin{figure}
    \centering
    \includegraphics[width = \linewidth]{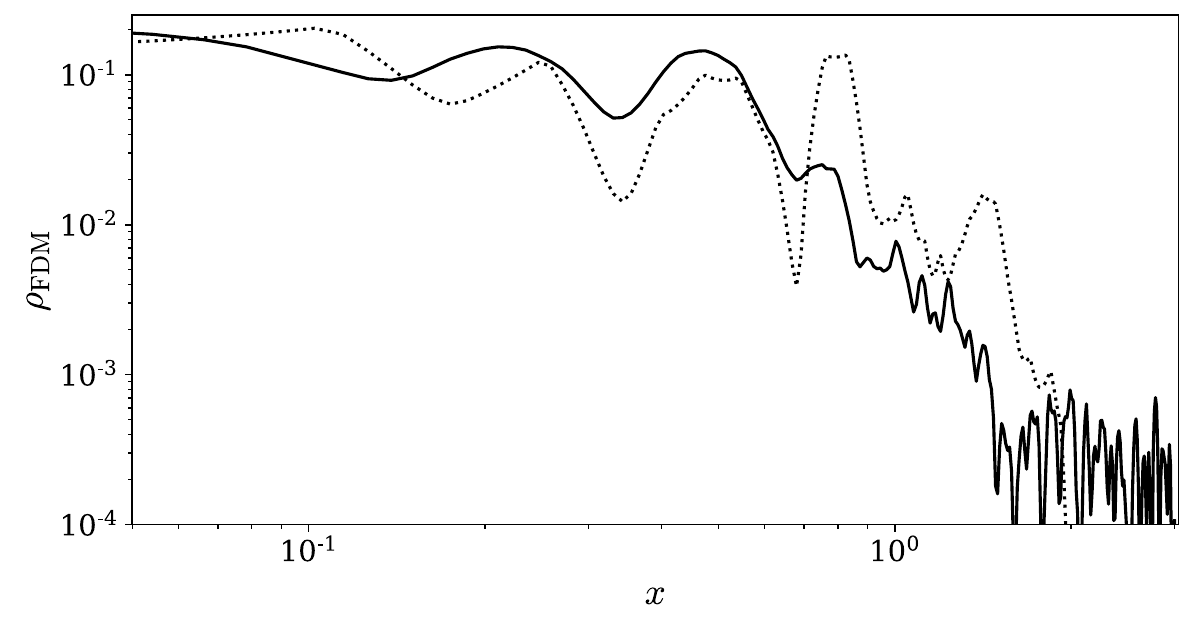}
    \caption{Density of the SF at the start (dotted) and end (solid) of the simulation. Averaged between the left and right side of the box.}
    \label{fig: FDM Density Evolution}
\end{figure}

In addition to the phase space plots of the SF seen in Fig. \ref{fig: Phasespace Evolution V2}, we plot the initial and final SF density in Fig. \ref{fig: FDM Density Evolution}. The characteristic size of the SF shrinks in both position and velocity space, from root-mean-square values $x_{\rm rms}\simeq 0.598$ and $v_{\rm rms} \simeq 0.823$ to $x_{\rm rms} \simeq 0.435$ and $v_{\rm rms} \simeq 0.569$. This demonstrates the losses of both potential and kinetic energy in the SF over the course of the simulation.

\subsection{Evolution of the half-mass radius and velocity dispersion}

As discussed in the introduction, \citet{Dalal_2022} argue that large regions of FDM parameter space are ruled out by the existence of dark matter dominated UFDs such as Segue 1 and Segue 2. The essence of their argument is that if the dark halos of these systems were composed of FDM with $mc^2\lesssim 3\times 10^{-19}\,{\rm eV}$, then the stellar dispersion $\sigma_\star$ and projected half-light radius $R_{1/2}$ would grow over time to the point of being inconsistent with observations. This argument is illustrated in their Figure 2, which shows $R_{1/2}$ and $\sigma_\star$ as functions of time for various values of $m$ and for different initial conditions for the stars. They find that both $R_{1/2}$ and $\sigma_\star^2$ roughly double over $\sim 10\,{\rm Gyr}$. Since these systems have dynamical times of order $1\,{\rm Myr}$, one can assume that approximate virial equilibrium is maintained. Thus, the virial mass one would infer from the stars, $M_{\rm vir}\simeq 4\sigma_\star^2 R_{1/2}$ \citep{wolf2010}, will have increased by a factor of four, which is consistent with having a stellar system that expands by a factor of two within a cuspy dark halo.

We now carry out a similar analysis in our star + SF/QP simulations. We first note that for our initial conditions, $\sigma_\star\simeq 0.842$ and $x_{1/2} \simeq 0.423$ where, by definition, half of the total surface density is contained in the region $|x|<x_{1/2}$. Therefore, one can define a virial surface density, in analogy with the virial mass, as $\Sigma_{\rm vir}\propto \sigma_*^2/x_{1/2}$. The constant of proportionality, which is obtained by setting $\Sigma_{\rm vir}=1/\pi$, the total surface density, and $\sigma_*$ and $x_{1/2}$ to their initial values, is $0.19$.

\begin{figure}
    \centering
    \includegraphics[width = \columnwidth]{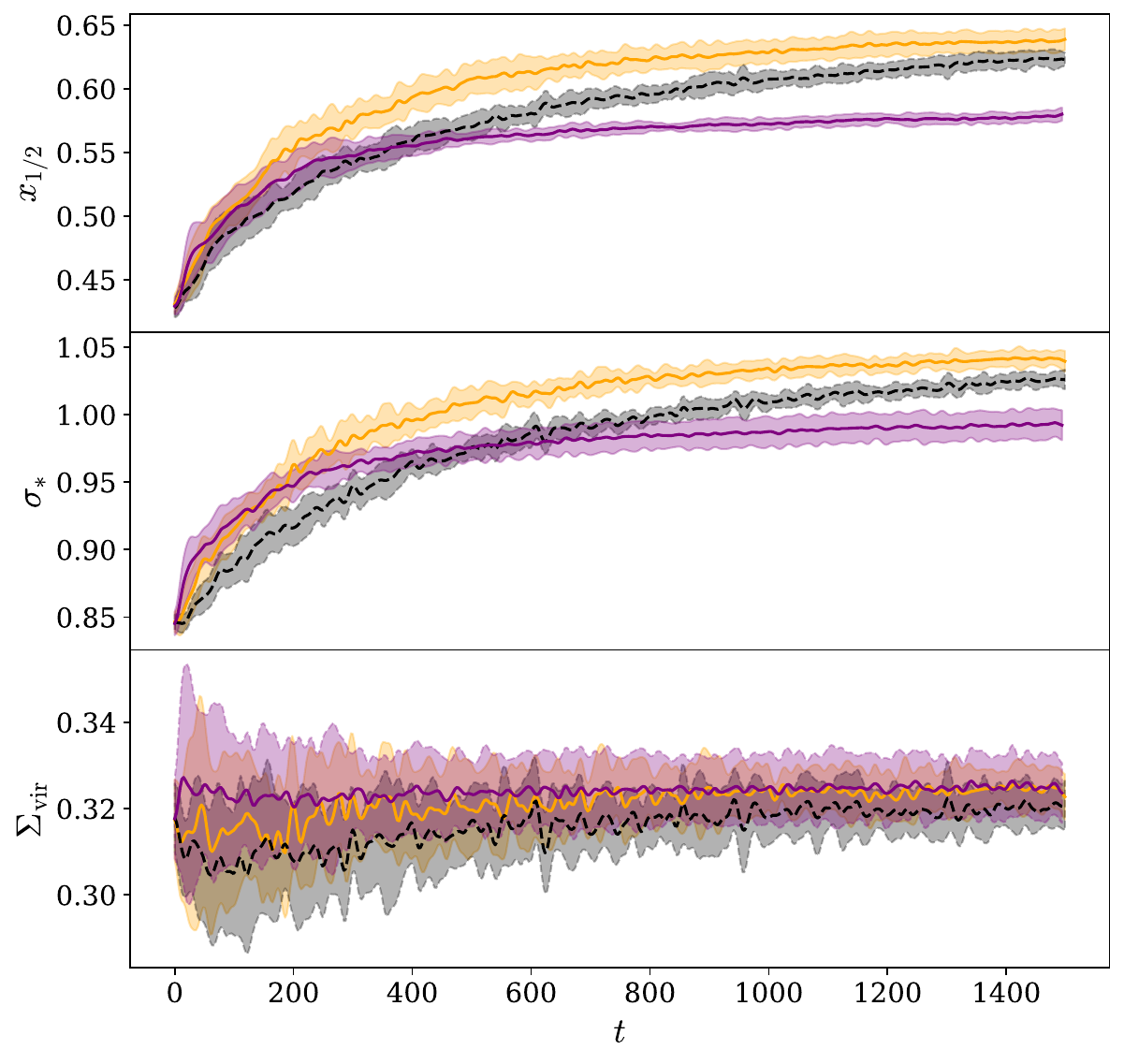}
    \caption{Half-mass radius (top) and velocity dispersion (bottom) of the particles over time in either regime. Orange and purple correspond to $5\times 10^4$ stars mixed with either QP or a SF, respectively. The dashed black curve corresponds to a simulation using double the number of QPs. Each curve is the average over 4 simulations. Similarly, the light bands represent the standard deviation across the 4 simulations for each regime. Note that,  for the sake of visibility, all curves have been smoothed using a Gaussian filter.}
    \label{fig: R and sigma}
\end{figure}

In Fig. \ref{fig: R and sigma} we compare the evolution of $x_{1/2}$ and $\sigma_\star$ for stars embedded in a SF halo and stars embedded in a halo of QPs. For each case, we run four simulations for $\simeq 375 t_{\rm dyn}$. We find that in the star+QP simulations, $x_{1/2}$ and $\sigma_\star$ increase steadily by $30\%$ and $20\%$, respectively. The virial surface density $\Sigma_{\rm vir}$ and hence the effective mass are roughly constant during the course of the simulation.

For $t\lesssim 50t_{\rm dyn}$ the evolution of $x_{1/2}$ and $\sigma_\star$ in the FDM and QP simulations are very similar. However, at later times, $x_{1/2}$ and $\sigma_\star$ are nearly constant in the stars + SF simulations and the overall increase in $x_{1/2}$ and $\sigma_\star^2$ is about half what it is in the stars + QPs simulations. To further explore the differences between QP and FDM simulations, we rerun the simulations with twice as many QPs. As expected, the change in $x_{1/2}$ and $\sigma_\star$ is decreased relative to the original QP simulation though it is still higher than the change found in the FDM simulations. Moreover, the heating at earlier times is of a lower rate.

\section{Summary and Discussion}\label{sec: Discussion}

Our aim in this paper has been to test the hypothesis that FDM can be treated as a system of QPs using standard N-body techniques. The main difference between QPs and the N-body systems normally encountered in cosmology and galactic dynamics is in how the QP masses are assigned. In our implementation of the quasiparticle hypothesis, QPs are uniformly distributed in phase space with a mass proportional to the DF at their initial location. 

Our results regarding the validity of the QP hypothesis are mixed. Isolated systems of dark matter exhibit small oscillations about an equilibrium state whether they are modeled by QPs or a SF. Furthermore, the amplitude of oscillations in the virial ratio and the frequency power spectra of these oscillations are very similar so long the relation between $N_{QP}$ and ${\cal R}$ given by equation \ref{eq:NscriptR} is satisfied. The main difference is in the force fluctuation power spectrum ${\cal P}_F(k)$ in that the SF power spectrum shows cut-off for scales below the de Broglie wavelength, which isn't present in the power spectra from QP simulations.

We also showed that FDM can dynamically heat stars whether it is described as a SF or system of QPs. In fact, at early times, the rate at which the half-mass radius and velocity dispersion increase is very similar. However, at later times, the QPs appear to be more efficient at heating the stars than the SF.

We contend that QPs, at least as we've modelled them, do best as a proxy for a SF in situations where the structure of the FDM component does not change appreciably with time. For example, QPs should provide a useful substitute for a SF in problems such as the disruption of stellar streams \citet{amorisco2018, Dalal_2021} or the heating of a disc in an FDM halo \citet{Hui_2017}. QPs can be reliably used in systems where dark matter dominates the potential and the stars can be treated as test particles, such as the analysis of UFDs in \citet{dutta2023} where the stars make a negligible contribution to the potential. On the other hand, there are many situations where the FDM distribution function changes significantly. We've given one example where the contributions to the potential from FDM and stars are comparable and where the FDM halo adiabatically condenses as the stellar system adiabatically expands. In these situations, the transfer of energy from a SF to stars may be significantly different than that for a system of QPs and the QP hypothesis might lead to erroneous conclusions. Of course, these conclusions were reached through one dimensional simulations. Though the essential physics of dynamical heating in one dimension is the same as in three dimensions, the details are very different. Ultimately, it will take three dimensional numerical experiments similar to the ones performed here to test the applicability of the QP hypothesis.

\section*{Acknowledgements}

We are grateful to Neal Dalal, Lam Hui, and Tomer Yavetz for useful conversations. We acknowledge the financial support of the Natural Sciences and Engineering Research Council of Canada. BZ also acknowledges the support of the McDonald Institute.

\section*{Data Availability}

The data underlying this article were generated by numerical calculations using original \textsc{Python} code written by the authors. The code incorporated routines from \textsc{NumPy} \citep{harris2020} and \textsc{SciPy} \citep{virtanen2020}. The statistical analysis in Section 3 was performed using the Markov chain Monte Carlo sampler \textsc{emcee} \citet{emcee2013}. Figure \ref{fig:cornerplot} was produced using the python package \textsc{corner.py} \citep{corner}. The data for the figures and the code will be shared on reasonable request to the authors.

\bibliographystyle{mnras}
\bibliography{FDMQP}  

\end{document}